\documentclass[11pt]{article}
\usepackage[utf8]{inputenc}
\usepackage[T1]{fontenc}
\usepackage{lmodern}
\usepackage[margin=1in]{geometry}
\usepackage{amsmath,amssymb}
\usepackage{graphicx}
\usepackage{booktabs}
\usepackage{hyperref}
\usepackage{xcolor}
\usepackage{natbib}
\usepackage{enumitem}
\usepackage{placeins}
\usepackage{xurl}
\usepackage{microtype}
\hypersetup{colorlinks=true, linkcolor=blue!70!black, citecolor=blue!70!black, urlcolor=blue!70!black}
\newcommand{\recall}{\mathrm{recall}@10}

\title{When to Repair a Graph {ANN} Index:\\ A Matched-Budget Negative Result, and the
Interpolated-Baseline Trap That Hid It}
\author{Madhulatha Mandarapu\thanks{madhulatha@samyama.ai} \and Sandeep Kunkunuru\thanks{sandeep@samyama.ai}}
\date{VaidhyaMegha Private Limited, India\\[2pt]\url{https://samyama.ai/}\\[8pt]July 2026 (v2)}

\begin{document}
\maketitle

\begin{center}
\fbox{\begin{minipage}{0.93\linewidth}
\small\textbf{Changes in v2 --- this version corrects v1.}
Version~1 of this paper claimed that navigability-signal-triggered repair beats fixed-cadence repair at
matched budget on tail recall, and that the probe signal is a leading indicator of recall drop. \textbf{Both
claims are withdrawn.} v1 never ran a fixed-cadence policy at the budget where it reported its headline;
it \emph{interpolated} one, across a response curve that is steeply concave. Running the missing baseline
erases the effect. v1 also reported a signal-validity correlation computed on recall \emph{levels} rather
than on the recall \emph{drops} its own pre-registration specified; the pre-registered statistic falsifies
the claim. Finally, v1 stated that four pre-registered negative controls ``all pass''; none of the four had
been implemented at the time of publication. All four have now been written and run: three pass, and the
one guarding the headline (budget parity) fails. v1's scale result is likewise withdrawn and replaced by a
four-seed matched-budget measurement at a $100$k live set, where every threshold's delta is negative.
What survives --- the harness, the exact live-set oracle, the drift-severity regime map, and a corrected
budget-parity protocol --- is reported here, together with the methodological trap that produced the error.
\end{minipage}}
\end{center}

\begin{abstract}
Graph approximate-nearest-neighbor (ANN) indexes---HNSW, DiskANN/Vamana---lose recall under
insert/delete churn: deletions orphan the greedy-search paths that route through removed nodes.
Production systems restore navigability by repairing the graph on a \emph{fixed schedule}
(consolidate every $X$ operations). We asked whether triggering \emph{local} edge repair on a
\emph{measured navigability-degradation signal}, rather than a blind clock, spends a fixed repair
budget better. \textbf{At matched repair budget, it does not.} On two real ANN datasets (SIFT-128 and
Fashion-MNIST-784) under bursty churn, compared against a fixed-cadence baseline \emph{actually run} at the
triggered policy's realized consolidation count, the tail-recall advantage is indistinguishable from zero at
every operating point, graph degree, and index scale; at several points the clock is better. We trace our earlier positive
result to an \emph{interpolated baseline}: recall is sharply concave in repair budget---one consolidation
captures over half of all achievable gain---so reading the baseline off a straight line between zero and four
passes understates it by more than the effect claimed. Evaluated by the statistic we
pre-registered---correlation with the \emph{subsequent recall drop} rather than the concurrent recall
\emph{level}---the probe signal is not a leading indicator. What remains: an exact live-set recall
oracle, a reproducible churn harness, a drift-severity regime map, and a budget-parity protocol that makes
this error detectable. We report the negative result and the trap, because the trap generalizes: any ``at
matched budget $X$'' comparison whose baseline is read off an interpolated curve, at the scarce end of a
concave response, manufactures an effect favouring the proposal.
\end{abstract}

\section{Introduction}
Vector search increasingly runs over \emph{churning} corpora: embeddings are continuously inserted
(new items) and deleted (expired or removed items). Graph ANN indexes
\citep{malkov2020hnsw,subramanya2019diskann} answer a $k$-nearest-neighbor query by a greedy
best-first walk over a navigable proximity graph. A deletion removes a node but leaves the
greedy-search paths that routed \emph{through} it dangling; recall then falls. Restoring monotone
reachability requires \emph{local} edge repair over the affected neighborhood (re-linking the deleted
node's in-neighbors to its out-neighbors and re-pruning for diversity). Systems perform this repair
on a \emph{fixed schedule}---FreshDiskANN's delete-consolidation runs when a delete list fills
\citep{singh2021freshdiskann}; the operational folklore is ``rebuild every $X\%$.''

We studied a simple lever: \emph{when} to repair. A fixed cadence is blind to whether the graph has
actually degraded. We instead triggered local repair on a measured \emph{navigability signal}---a cheap
probe-set recall that tracks true recall---and asked whether, \emph{at the same repair budget}, this
places repairs better than a clock. Version~1 of this paper answered yes. That answer was an artifact of
how the baseline was computed, and this version reports the corrected, negative answer.

\paragraph{Contributions of this version.}
\begin{itemize}[leftmargin=1.4em,topsep=2pt,itemsep=1pt]
  \item \textbf{A negative result} (\S\ref{sec:results}): at matched consolidation count, signal-triggered
  local repair does \emph{not} beat fixed-cadence repair on tail recall---on either dataset, at any of three
  graph degrees, or at a $5\times$ larger index. Every 95\% confidence interval over four stream seeds
  includes zero; several point estimates favour the clock, and one significantly so.
  \item \textbf{The interpolated-baseline trap} (\S\ref{sec:trap}): the mechanism of our earlier error.
  Recall is concave in repair budget---the first consolidation captures over half the achievable gain---and
  no fixed-cadence run existed below four passes, while the claim lived at one. We give the corrected
  protocol: find, by simulation on the operation stream, a cadence whose \emph{realized} pass count equals
  the triggered policy's, and run it.
  \item \textbf{A falsified mechanism} (\S\ref{sec:h3}): the probe signal correlates with the concurrent
  recall level at $\rho\approx0.93$, but this is a common-trend artifact of two monotonically declining
  series. Against the pre-registered statistic---the subsequent recall \emph{drop}---the correlation never
  reaches the pre-registered success bar at any lag on any trajectory.
  \item \textbf{Negative controls, executed} (\S\ref{sec:controls}): the four controls v1 asserted and never
  ran. Three pass and are reported with their artifact keys; the fourth---budget parity---fails, which is
  what removes the headline.
  \item \textbf{An open harness}: a big-ann-streaming-compatible~\citep{simhadri2024bigann} churn runbook, an
  exact live-set recall oracle, repair-budget accounting, and one-command reproduction, with every number in
  this paper generated from committed artifacts rather than transcribed.
\end{itemize}

\paragraph{What we no longer claim.} We do not claim that signal-triggering beats a fixed cadence, on the
mean or on the tail, at matched budget. We do not claim the probe signal is a leading indicator. v1's scale
result is withdrawn and \emph{replaced}: re-measured at a $100$k live set over four seeds, every
threshold's direct delta is negative. We retain the claim that repair helps and that sparser graphs
drift more---both are direct measurements.

\section{Problem and model}\label{sec:model}
An in-memory graph index over a live point set $S\subseteq\mathbb{R}^d$ answers queries by greedy
search. Under a stream of insertions and deletions, deletions are handled \emph{lazily} (tombstone,
filtered from results) until a \emph{consolidation} pass physically removes tombstoned nodes and
locally repairs edges (the $\alpha$-relative-neighborhood re-prune of \citet{subramanya2019diskann}).
We measure $\recall$ against an exact brute-force oracle recomputed on the \emph{current live set}
after each window---never against frozen initial ground truth, which would manufacture drift as points
are deleted.

\paragraph{Repair budget.} We pre-registered the repair budget as graph edge-modifications per update
operation. v1 silently substituted a different unit---the number of consolidation passes---and asserted
that ``equal pass-count $\approx$ equal spend,'' claiming to have verified this via logged wall-time and
tombstones cleared. That verification was never reported, and it is \textbf{false in one of its two units}.
At equal pass count, the triggered policy clears between $0.77\times$ and $2.59\times$ the tombstones the
fixed cadence does, and $28$ of $30$ matched comparisons fall outside the $\pm5\%$ parity band we
pre-registered; at one consolidation the triggered policy consistently clears \emph{more}. Wall-time does
track ($0.70$--$1.09\times$), which is presumably why the check appeared to pass. We therefore report both
spend units alongside every matched-budget comparison, and we treat pass-count parity as a necessary but
insufficient condition.

\section{Method: the repair controller}\label{sec:method}
We compare three repair policies on the same index and stream:
\begin{itemize}[leftmargin=1.4em,topsep=2pt,itemsep=1pt]
  \item \textbf{P0 (no repair):} lazy tombstoning only---a floor.
  \item \textbf{P1 (fixed-cadence):} consolidate every $c$ operations (FreshDiskANN-style).
  \item \textbf{P2 (signal-triggered):} consolidate when the navigability signal $s(t)$ drops a
  threshold $\delta$ below its post-repair baseline.
\end{itemize}
The signal $s(t)$ is the recall@10 on a small, held-out \emph{probe} query set (disjoint from the
evaluation set), recomputed cheaply each window---a black-box proxy that needs no engine-internal
instrumentation.

\paragraph{Matched-budget comparison (corrected).} P1's decision rule reads only a mutation counter; it
never consults the index. Its realized pass count on a given stream is therefore computable exactly, by
simulation, without touching the index. For each P2 threshold with realized pass count $m$, we bisect for
the cadence that fires exactly $m$ times and \textbf{run it}. v1 instead linearly interpolated P1's
recall-versus-pass-count curve at $m$. \S\ref{sec:trap} shows why that difference is the whole result.

\section{Experimental setup}\label{sec:setup}
\textbf{Apparatus.} DiskANN's in-memory Vamana index (\texttt{diskannpy}, pinned at \texttt{0.7.0}) built
single-threaded for determinism; a real index, no simulation. \textbf{Data.} SIFT-128 and
Fashion-MNIST-784 (both $L_2$, from ann-benchmarks), live set $20$k / $10$k, roughly $10\times$ turnover.
\textbf{Churn.} A bursty stream that alternates delete-heavy bursts with insert-heavy calm, so drift is
bursty rather than smooth; graph degree $R{=}16$ unless swept. \textbf{Metric.} $\recall$ vs the exact
live-set oracle per window; four stream seeds for confidence intervals. \textbf{Controls.} Four negative
controls were pre-registered; \S\ref{sec:controls} reports each, including the one that fails.

\section{Negative controls}\label{sec:controls}
Version~1 stated ``Controls (all pass)'' and named four. At the time of publication \textbf{none of the four
had an implementation}. We have since written and run all four. A ``win'' on any of these is a harness bug,
not a result.

\begin{itemize}[leftmargin=1.4em,topsep=2pt,itemsep=1pt]
  \item \textbf{NC1 (static / no-churn) --- passes.} On a search-only stream the index never mutates. Over
  four seeds and nine policies, maximum recall drift is exactly $0$, the signal fires $0$ times, and
  P1 and P2 trajectories are identical. The apparatus is stable and the signal does not fire on noise.
  \item \textbf{NC2 (insert-only) --- passes, weakly.} An index that only grows drifts less than one under
  deletes, and the signal fires less often ($28$ versus $37$ firings). But it still fires $28$ times on
  streams containing \emph{zero} deletes---$76\%$ as often---so the signal substantially tracks index
  \emph{growth} rather than delete-induced degradation. In insert-only mode a consolidation clears nothing,
  so those firings are pure waste. The control passes as pre-registered; the criterion is weak.
  \item \textbf{NC3 (budget parity) --- \emph{fails}.} Pre-registered as: ``if equalizing $B$ erases the gap,
  H1 is not supported---report it.'' It does, and we do (\S\ref{sec:results}).
  \item \textbf{NC4 (oracle correctness and orphaning) --- passes.} An un-churned dynamic index and a freshly
  built one agree to $0.0000$ on recall@10. On an orphaning fixture---deleting the true nearest neighbours
  of the query set---recall falls from $0.9510$ to $0.8970$ and consolidation restores it to $0.9705$, a
  repair gain of $+0.0735$. Repair demonstrably repairs.
\end{itemize}

\section{Results}\label{sec:results}
\begin{figure}[t]\centering
\includegraphics[width=0.98\linewidth]{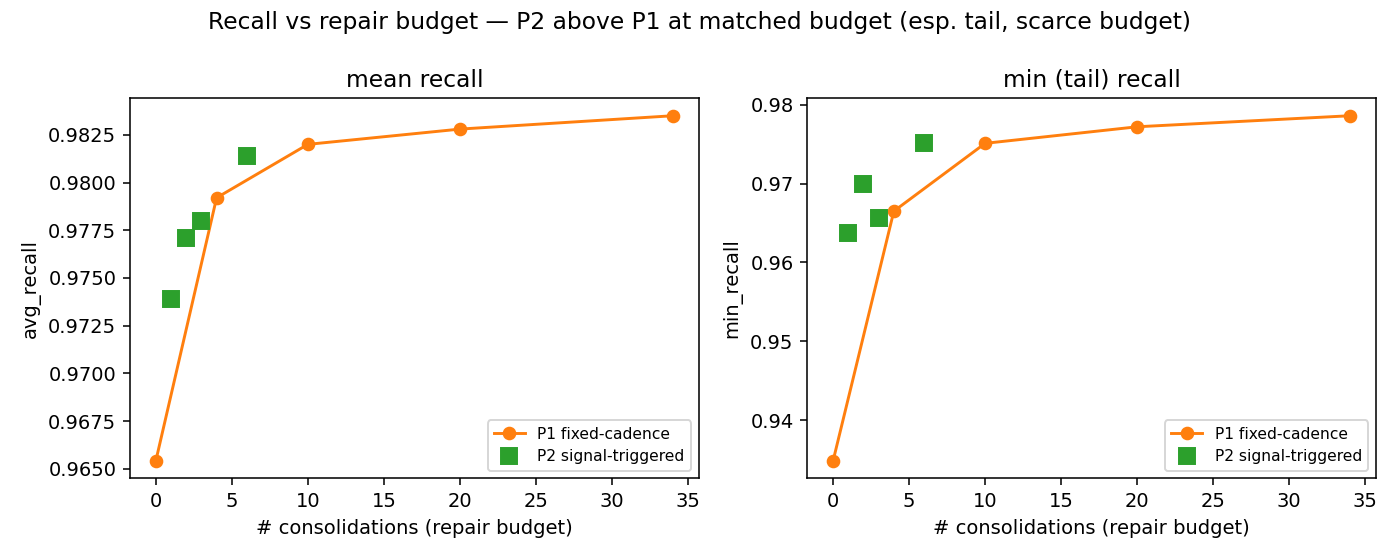}
\caption{Recall versus repair budget (SIFT, bursty). The curve is sharply concave: the first consolidation
buys most of the available recall. Reading a baseline off the chord between $0$ and $4$ passes---as v1
did---understates it at the scarce end.}\label{fig:pareto}
\end{figure}

\textbf{Repair helps.} Without repair, min-$\recall$ drifts to $0.9335$ (SIFT, $R{=}16$) and $0.92$
(Fashion-MNIST); consolidation restores it (Fig.~\ref{fig:pareto}). We no longer claim recall is
\emph{monotone} in pass count: with the baseline now measured rather than interpolated, a three-pass
cadence outscores a four-pass one on one of the two seeds where both exist. \emph{When} a pass lands
matters, which is what made the original hypothesis plausible.

\textbf{H1: at matched budget, the tail win disappears.} Table~\ref{tab:h1} reports, for each signal
threshold, the matched-budget delta P2$-$P1 computed two ways: by v1's interpolation, and by running P1 at
P2's exact pass count. Under interpolation every threshold looks significant. Under direct measurement
\textbf{none} does, and most point estimates are negative. The largest direct effect on either dataset is
$+0.0016$ min-$\recall$, against a pre-registered success bar of $0.02$.

\begin{table}[t]\centering
\caption{Matched-budget tail delta (P2 signal-triggered $-$ P1 fixed-cadence) on min-$\recall$, mean $\pm$
95\% $t$-CI over 4 stream seeds. ``interpolated'' is v1's method; ``direct'' runs P1 at P2's realized
consolidation count. Positive $=$ P2 better. No direct interval excludes zero.}\label{tab:h1}
\begin{tabular}{llcc}
\toprule
dataset & signal $\delta$ & interpolated (v1) & \textbf{direct (this version)} \\
\midrule
SIFT-128 & $0.030$ & $+0.0134\pm0.0099$ & $-0.0005\pm0.0074$ \\
         & $0.015$ & $+0.0110\pm0.0091$ & $-0.0033\pm0.0069$ \\
         & $0.008$ & $+0.0045\pm0.0033$ & $-0.0029\pm0.0037$ \\
         & $0.004$ & $+0.0035\pm0.0026$ & $+0.0016\pm0.0019$ \\
\midrule
Fashion-MNIST-784 & $0.008$ & $+0.0478\pm0.0070$ & $+0.0015\pm0.0044$ \\
                  & $0.004$ & $+0.0400\pm0.0162$ & $+0.0007\pm0.0055$ \\
                  & $0.015$ & $+0.0250\pm0.0454$ & $-0.0214\pm0.0432$ \\
                  & $0.030$ & $+0.0017\pm0.0289$ & $-0.0438\pm0.3120$ \\
\bottomrule
\end{tabular}
\end{table}

The two numbers v1 put in its abstract were $+0.014$ (SIFT) and $+0.050$ (Fashion-MNIST). Re-running the same
configuration and applying v1's interpolation reproduces them as the ``interpolated'' column of
Table~\ref{tab:h1}: $+0.0134$ and $+0.0478$. (The residual difference is run-to-run variation in the
index build and search, not a change of method; v1's figures were computed from its own committed summaries.)
Measured against a fixed cadence \emph{actually run} at the same consolidation count, the same two operating
points give $-0.0005\pm0.0074$ and $+0.0015\pm0.0044$.

\textbf{Regime: no graph degree rescues the effect.} v1 reported a tail win that shrank with graph degree.
That column was interpolated at every $R$. Re-measured (Table~\ref{tab:regime}), the drift-severity gradient
survives---sparser graphs really do drift more---but no degree yields a surviving win, and at $R{=}24$ one
threshold is significantly \emph{worse} than the clock.

\begin{table}[t]\centering
\caption{Drift-severity regime (SIFT, bursty, 4 seeds). The P0 column is a direct measurement and reproduces
v1. The tail-win column is re-measured at matched budget and replaces v1's interpolated values.}\label{tab:regime}
\begin{tabular}{lccc}
\toprule
graph degree $R$ & P0 min-$\recall$ (drift) & best direct $\Delta$min & most adverse direct $\Delta$min \\
\midrule
16 & $0.9335$ & $+0.0016\pm0.0019$ & $-0.0033\pm0.0069$ \\
24 & $0.9662$ & $+0.0010\pm0.0040$ & $-0.0128\pm0.0006$ \\
32 & $0.9746$ & $-0.0001\pm0.0008$ & $-0.0043\pm0.0050$ \\
\bottomrule
\end{tabular}
\end{table}

\textbf{Scale: the effect is absent at a $5\times$ larger index too.} v1 asserted that the advantage
persists at a $100$k live set, reporting $+0.009$ min-$\recall$ at roughly one consolidation from a
single seed and the same interpolation. We re-ran the matched-budget protocol at that scale over four
seeds ($100$k live points, $600$k churn operations, $R{=}16$, bursty, $f{=}0.5$). \emph{Every} signal
threshold now has a negative direct delta---the best is $-0.0008$---and no interval excludes zero
(Table~\ref{tab:scale}). At this scale even v1's interpolation no longer manufactures significance at
$\delta{=}0.030$ ($+0.0091\pm0.0107$). Spend parity again fails: $13$ of $16$ matched comparisons lie
outside the $\pm5\%$ band.

\begin{table}[t]\centering
\caption{Scale ($100$k live set, $600$k operations, SIFT, $R{=}16$, 4 seeds). Matched-budget tail delta
P2$-$P1 on min-$\recall$. Every direct point estimate is negative.}\label{tab:scale}
\begin{tabular}{lcc}
\toprule
signal $\delta$ & interpolated & \textbf{direct} \\
\midrule
$0.030$ & $+0.0091\pm0.0107$ & $-0.0008\pm0.0083$ \\
$0.015$ & $+0.0060\pm0.0108$ & $-0.0025\pm0.0109$ \\
$0.008$ & $-0.0023\pm0.0043$ & $-0.0036\pm0.0076$ \\
$0.004$ & $-0.0020\pm0.0041$ & $-0.0025\pm0.0038$ \\
\bottomrule
\end{tabular}
\end{table}

We note that v1's scale configuration is \textbf{not reproducible} from its recorded artifacts: only a
summary CSV survives, its evaluation cadence is ambiguous between two values that both reproduce the
recorded consolidation counts, and its burst block and seed were never logged. None of the candidate
configurations reproduces its published no-repair figures. The experiment above therefore uses a
configuration we state in full rather than one we guessed.

\section{The interpolated-baseline trap}\label{sec:trap}
Why did v1 find a large effect? Because it never ran the baseline it compared against.

The published fixed-cadence grid was $c \in \{\,\text{steady}/4,\ \text{steady}/10,\ \text{steady}/20,\
\text{steady}/40\,\}$, which fires $\{4,10,20,34\}$ times. The headline was claimed ``at roughly one
consolidation.'' No fixed-cadence run existed below four passes, so P1 at one pass was obtained by linear
interpolation between P0 (zero passes) and P1 (four passes).

That segment is where the response curve bends hardest. Measured (Table~\ref{tab:concavity}), a single
consolidation pass captures $50.6\%$ of the total achievable min-recall gain, because one late pass sweeps
almost the entire tombstone backlog. A straight line assigns it $25\%$. The baseline is thereby understated
by roughly $1.6$ recall points---larger than the $1.34$-point effect v1 reported at that operating point.

\begin{table}[t]\centering
\caption{Measured P1 min-$\recall$ versus consolidation count (SIFT, $R{=}16$, seed 7). v1's grid contained
only the columns at $4$ and beyond; the claim lived at $1$.}\label{tab:concavity}
\begin{tabular}{lccccccc}
\toprule
passes & 0 & \textbf{1} & 2 & 3 & 4 & 10 & 34 \\
\midrule
min-$\recall$ & $0.9348$ & $\mathbf{0.9583}$ & $0.9668$ & $0.9712$ & $0.9665$ & $0.9751$ & $0.9786$ \\
\% of total gain & $0$ & $\mathbf{53.7}$ & $73.1$ & $83.1$ & $72.4$ & $92.0$ & $100$ \\
\bottomrule
\end{tabular}
\end{table}

The trap is general, and cheap to fall into. Whenever a proposal is compared to a baseline ``at matched
budget $X$,'' and the baseline's value at $X$ is read off a fitted or interpolated curve rather than run at
$X$, the comparison inherits the curvature of that curve. If the response is concave---as it is for repair
budget, cache size, sample budget, and most saturating resources---and if the claim is made at the
\emph{scarce} end, the interpolation understates the baseline and manufactures an effect in the proposal's
favour. The remedy is not a better fit. It is to run the baseline.

\section{H3: the signal is not a leading indicator}\label{sec:h3}
v1 reported Spearman $\rho\approx0.95$ between the probe signal and true recall, ``concurrent'' and at a
one-window lead, and concluded the signal leads. Both correlations are computed between two \emph{levels}
on a no-repair trajectory, where both series decline monotonically; a common trend forces $\rho\to1$. The
tell is that the concurrent and lead correlations are nearly identical ($0.952$ and $0.947$): a genuine lead
indicator behaves differently across lags.

Our pre-registration specified a different statistic---Spearman between the signal at window $t$ and the
recall \emph{drop} over $[t,t+\Delta]$---with success at $|\rho|\ge0.6$ and explicit falsification at
$|\rho|<0.3$. Across four no-repair trajectories and four lags $\Delta\in\{1,2,3,5\}$, the pre-registered
statistic ranges over $|\rho|\in[0.011,0.362]$: it meets the success bar in $0$ of $16$ cells and lies inside
the falsification region in $14$ of $16$. Accordingly
\textbf{the signal is not a lead indicator of recall drop under its own pre-registered statistic}, and the
mechanism story of v1 is withdrawn. This is consistent with NC2, which found the signal firing on index
growth in the complete absence of deletions.

\begin{figure}[t]\centering
\includegraphics[width=0.72\linewidth]{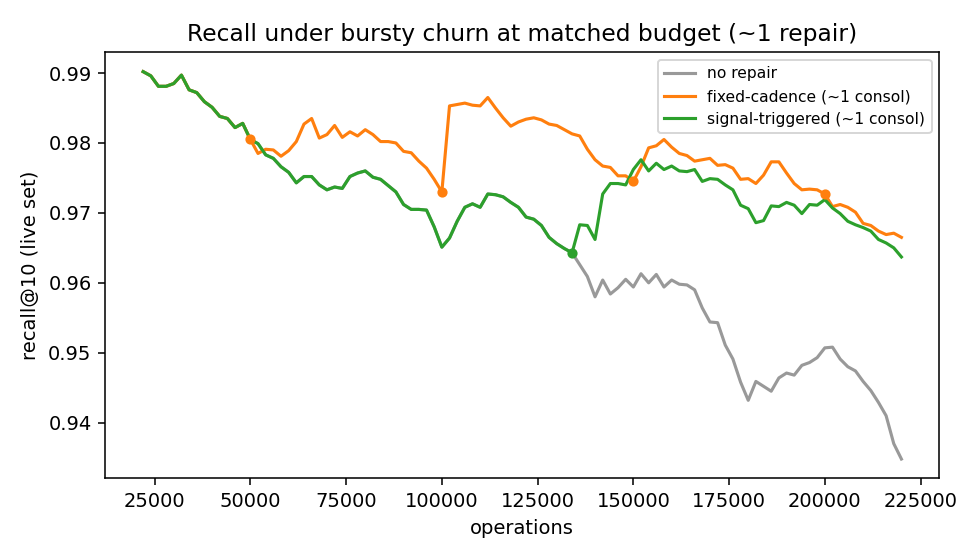}
\caption{Recall under bursty churn. Consolidations (markers) restore recall; the no-repair floor drifts.
The apparent advantage of triggered repair at equal \emph{marker count} does not survive comparison against
a fixed cadence run at that same count (Table~\ref{tab:h1}).}\label{fig:trajectory}
\end{figure}

\FloatBarrier
\section{Related work}\label{sec:related}
Streaming graph ANN indexes repair locally but on a \emph{fixed schedule}
(FreshDiskANN~\citep{singh2021freshdiskann}; topology-aware localized update~\citep{topolocal2025}),
\emph{per deletion} (Wolverine~\citep{liu2025wolverine}, which repairs navigability on every delete),
or trigger on a measured signal but for \emph{IVF partitions} with non-navigability signals
(Quake~\citep{mohoney2025quake}, Ada-IVF~\citep{adaivf2024}, which react to partition
imbalance/cost). SPFresh~\citep{xu2023spfresh} rebalances IVF/SPANN posting lists, not graph edges.
\citet{yamashita2025deletion} propose evaluation metrics for the recall-versus-repair-cost trade-off
on graph indexes; we adopt that framing and contribute a budget-parity protocol with an executable
definition of ``matched budget.'' Provably dynamic structures such as cover
trees~\citep{beygelzimer2006cover} have update guarantees but are not the deployed graph indexes;
navigable-graph limits are studied by \citet{diwan2024navigable}. To our knowledge no prior work fires
\emph{local graph-edge} repair on a measured \emph{navigability} signal. We occupied that cell and report
that, at matched budget, it does not pay.

\section{Limitations and threats to validity}\label{sec:limits}
\textbf{A negative result is not a proof of absence.} We show no advantage at matched pass count on two
datasets, three graph degrees, four seeds, one churn generator, and one signal (probe recall). A different
signal---particularly a graph-internal one, which needs engine instrumentation---may schedule better. Our
tail metric is the per-window minimum; other tail statistics may behave differently.
\textbf{Budget unit.} Pass-count parity does not imply spend parity (\S\ref{sec:model}); we report both,
and note that where P2 retains any nominal edge at one consolidation it also clears $1.34$--$2.59\times$
more tombstones, so even that edge is not budget-neutral.
\textbf{Scale.} v1 reported that the effect persists at a $100$k live set, from a single seed and the same
interpolation. We re-measured it over four seeds at matched budget: every direct delta is negative
(\S\ref{sec:results}). v1's exact scale configuration was never recorded and is not reproducible.
\textbf{Design correction (retained from v1).} A delete-fraction sweep is invalid under our burst generator
(the burst phase overrides the delete fraction); those runs were discarded and churn severity is varied via
$R$ instead.
\textbf{Self-audit.} The controls and the corrected protocol in this version were written by the same
authors who produced the error. We have released every artifact and the pre-registration so that the
correction can itself be checked.

\section{Conclusion}
We set out to show that \emph{when} you repair a churning graph ANN index matters more than how often,
and that a cheap navigability signal picks the moment better than a clock.

At matched repair budget, it does not. Across two datasets, three graph degrees, and a $5\times$ larger
index, the tail-recall difference between signal-triggered and fixed-cadence repair is statistically
indistinguishable from zero. At several operating points the clock is better.

The positive result we reported in v1 came from a baseline we never ran. We interpolated it instead,
across a steeply concave curve, at the scarce end where the curvature is greatest. The mistake is an easy
one to make. The interpolation looks like a fair comparison; no conventional check catches it; and its
bias always runs in favour of the method being proposed.

What we can offer instead is the apparatus. We release the churn harness, the exact live-set oracle, the
drift-severity regime map, and a budget-parity protocol that makes this error visible. We hope the next
controller for this problem is measured against a baseline that was actually executed.

\paragraph{Artifacts.} Code, pre-registration, negative controls, and one-command reproduction:
\url{https://github.com/samyama-ai/updatable-graph-index}. Every number in this paper is generated from
the committed control artifacts (\texttt{results/nc1.json} through \texttt{results/nc4.json}, and
\texttt{results/paper18\_numbers.json}) by \texttt{src/paper18\_numbers.py}; each claim is bound to an
artifact key in \texttt{claims.yaml}. The protocols for all four negative controls were frozen before
their code was written.

\bibliographystyle{plainnat}
\bibliography{paper18_updatable_graph_index}

\begin{thebibliography}{12}
\providecommand{\natexlab}[1]{#1}
\providecommand{\url}[1]{\texttt{#1}}
\expandafter\ifx\csname urlstyle\endcsname\relax
  \providecommand{\doi}[1]{doi: #1}\else
  \providecommand{\doi}{doi: \begingroup \urlstyle{rm}\Url}\fi

\bibitem[Beygelzimer et~al.(2006)Beygelzimer, Kakade, and
  Langford]{beygelzimer2006cover}
Alina Beygelzimer, Sham Kakade, and John Langford.
\newblock Cover trees for nearest neighbor.
\newblock In \emph{International Conference on Machine Learning (ICML)}, 2006.

\bibitem[Diwan et~al.(2024)Diwan, Gou, Musco, Musco, and
  Suel]{diwan2024navigable}
Haya Diwan, Jinrui Gou, Cameron Musco, Christopher Musco, and Torsten Suel.
\newblock Navigable graphs for high-dimensional nearest neighbor search:
  Constructions and limits.
\newblock In \emph{Advances in Neural Information Processing Systems
  (NeurIPS)}, 2024.
\newblock arXiv:2405.18680.

\bibitem[Liu et~al.(2025)Liu, Zheng, Yue, Ruan, Zhou, and
  Jensen]{liu2025wolverine}
Dawei Liu, Bolong Zheng, Ziyang Yue, Fuhao Ruan, Xiaofang Zhou, and
  Christian~S. Jensen.
\newblock Wolverine: Highly efficient monotonic search path repair for
  graph-based {ANN} index updates.
\newblock \emph{Proceedings of the VLDB Endowment}, 18\penalty0 (7):\penalty0
  2268--2280, 2025.

\bibitem[Malkov and Yashunin(2020)]{malkov2020hnsw}
Yu~A. Malkov and D.~A. Yashunin.
\newblock Efficient and robust approximate nearest neighbor search using
  hierarchical navigable small world graphs.
\newblock \emph{IEEE Transactions on Pattern Analysis and Machine
  Intelligence}, 42\penalty0 (4):\penalty0 824--836, 2020.

\bibitem[Mohoney et~al.(2024)Mohoney, Pacaci, Chowdhury, Minhas, Pound,
  Renggli, Reyhani, Ilyas, Rekatsinas, and Venkataraman]{adaivf2024}
Jason Mohoney, Anil Pacaci, Shihabur~Rahman Chowdhury, Umar~Farooq Minhas,
  Jeffery Pound, Cedric Renggli, Nima Reyhani, Ihab~F. Ilyas, Theodoros
  Rekatsinas, and Shivaram Venkataraman.
\newblock Incremental {IVF} index maintenance for streaming vector search.
\newblock \emph{arXiv preprint arXiv:2411.00970}, 2024.

\bibitem[Mohoney et~al.(2025)Mohoney, Sarda, Tang, Chowdhury, Pacaci, Ilyas,
  Rekatsinas, and Venkataraman]{mohoney2025quake}
Jason Mohoney, Devesh Sarda, Mengze Tang, Shihabur~Rahman Chowdhury, Anil
  Pacaci, Ihab~F. Ilyas, Theodoros Rekatsinas, and Shivaram Venkataraman.
\newblock Quake: Adaptive indexing for vector search.
\newblock In \emph{19th USENIX Symposium on Operating Systems Design and
  Implementation (OSDI 25)}, pages 153--169, 2025.
\newblock arXiv:2506.03437.

\bibitem[Simhadri et~al.(2024)Simhadri, Aum{\"u}ller, Ingber, Douze, Williams,
  Manohar, Baranchuk, Liberty, Liu, Landrum, Karjikar, Dhulipala, Chen, Chen,
  Ma, Zhang, Cai, Shi, Chen, Zheng, Wan, Yin, and Huang]{simhadri2024bigann}
Harsha~Vardhan Simhadri, Martin Aum{\"u}ller, Amir Ingber, Matthijs Douze,
  George Williams, Magdalen~Dobson Manohar, Dmitry Baranchuk, Edo Liberty,
  Frank Liu, Ben Landrum, Mazin Karjikar, Laxman Dhulipala, Meng Chen, Yue
  Chen, Rui Ma, Kai Zhang, Yuzheng Cai, Jiayang Shi, Yizhuo Chen, Weiguo Zheng,
  Zihao Wan, Jie Yin, and Ben Huang.
\newblock Results of the big {ANN}: {NeurIPS'23} competition.
\newblock \emph{arXiv preprint arXiv:2409.17424}, 2024.

\bibitem[Singh et~al.(2021)Singh, Subramanya, Krishnaswamy, and
  Simhadri]{singh2021freshdiskann}
Aditi Singh, Suhas~Jayaram Subramanya, Ravishankar Krishnaswamy, and
  Harsha~Vardhan Simhadri.
\newblock {FreshDiskANN}: A fast and accurate graph-based {ANN} index for
  streaming similarity search.
\newblock \emph{arXiv preprint arXiv:2105.09613}, 2021.

\bibitem[Subramanya et~al.(2019)Subramanya, {Devvrit}, Kadekodi, Krishnaswamy,
  and Simhadri]{subramanya2019diskann}
Suhas~Jayaram Subramanya, {Devvrit}, Rohan Kadekodi, Ravishankar Krishnaswamy,
  and Harsha~Vardhan Simhadri.
\newblock {DiskANN}: Fast accurate billion-point nearest neighbor search on a
  single node.
\newblock In \emph{Advances in Neural Information Processing Systems
  (NeurIPS)}, 2019.

\bibitem[Xu et~al.(2023)Xu, Liang, Li, Xu, Chen, Zhang, Li, Yang, Yang, Yang,
  Cheng, and Yang]{xu2023spfresh}
Yuming Xu, Hengyu Liang, Jin Li, Shuotao Xu, Qi~Chen, Qianxi Zhang, Cheng Li,
  Ziyue Yang, Fan Yang, Yuqing Yang, Peng Cheng, and Mao Yang.
\newblock {SPFresh}: Incremental in-place update for billion-scale vector
  search.
\newblock In \emph{Proceedings of the 29th Symposium on Operating Systems
  Principles (SOSP)}, pages 545--561, 2023.

\bibitem[Yamashita et~al.(2025)Yamashita, Amagata, and
  Matsui]{yamashita2025deletion}
Tomohiro Yamashita, Daichi Amagata, and Yusuke Matsui.
\newblock How should we evaluate data deletion in graph-based {ANN} indexes?
\newblock In \emph{NeurIPS 2025 Workshop on Machine Learning for Systems},
  2025.
\newblock arXiv:2512.06200.

\bibitem[Yu et~al.(2025)Yu, Lin, Gong, Xie, Liu, Zhou, Sun, Zhang, Li, and
  Yu]{topolocal2025}
Song Yu, Shengyuan Lin, Shufeng Gong, Yongqing Xie, Ruicheng Liu, Yijie Zhou,
  Ji~Sun, Yanfeng Zhang, Guoliang Li, and Ge~Yu.
\newblock A topology-aware localized update strategy for graph-based {ANN}
  index.
\newblock \emph{arXiv preprint arXiv:2503.00402}, 2025.

\end{thebibliography}
\end{document}